\documentclass[a4paper,11pt]{article}
\usepackage[compact]{titlesec}
\titlespacing{\section}{0pt}{2ex}{1ex}
\titlespacing{\subsection}{0pt}{1ex}{0ex}
\titlespacing{\subsubsection}{0pt}{0.5ex}{0ex}

\usepackage[scriptsize]{caption}
\usepackage[margin=1in]{geometry}
\usepackage[bookmarks, hyperindex, pdftex]{hyperref}
\usepackage[american]{babel} 
\usepackage{balance}
\usepackage{amsmath} 
\usepackage{multirow}
\usepackage{url}
\usepackage{amsthm}
\usepackage{amssymb}
\usepackage{graphicx} 
\usepackage{subfig} 
\usepackage{booktabs} 
\usepackage{url} 
\usepackage{paralist} 
\usepackage{cite}
\usepackage{authblk}
\usepackage{framed}

\usepackage[utf8]{inputenc}

\usepackage{tikz}
\usetikzlibrary{arrows}

\usepackage[numbers,sort&compress,square]{natbib} 

\newcommand{\COMPNAME}{\texttt{BC-ZIP}}

   {\vspace*{-4pt}
    \begin{enumerate}\itemsep=0pt}%
   {\end{enumerate}
    \vspace*{-2pt}}

   {\vspace*{-4pt}
    \begin{itemize}\itemsep=0pt}%
   {\end{itemize}
    \vspace*{-2pt}}

\newcommand{\LZSE}{\texttt{LZ77}}

\newcommand{\LZF}{\texttt{LZ4}}
\newcommand{\G}{\ensuremath{\mathcal{G}}}
\newcommand{\Scosts}{\ensuremath{s_\text{costs}}}
\newcommand{\Tcosts}{\ensuremath{t_\text{costs}}}

\newcommand{\LZP}[2]{\ensuremath{\langle #1, #2 \rangle}}
\newcommand{\LZC}[1]{\ensuremath{\langle 0, #1 \rangle}}

\newcommand{\tmax}{\ensuremath{t_{\text{max}}}}
\newcommand{\smax}{\ensuremath{s_{\text{max}}}}
\newcommand{\miss}[1]{\ensuremath{t_{#1}}}

\newcommand{\paths}[3]{\ensuremath{\operatorname{ps}\left(#1, #2, #3\right)}}
\newcommand{\Txt}{\ensuremath{{\cal S}}}
\newcommand{\Pre}{\ensuremath{{\sf Pref}}}
\newcommand{\Su}{{\ensuremath{\sf Suf}}}

\newcounter{thm}

\newtheorem{Theorem}[thm]{Theorem}

\newtheorem{Definition}[thm]{Definition}
\newtheorem{Lemma}[thm]{Lemma}
\newtheorem{Fact}[thm]{Fact}
\newtheorem{Property}[thm]{Property}
\newtheorem{Corollary}[thm]{Corollary}

\newcounter{save}

\newcommand\defer[4]{
  \expandafter\edef\csname deflemnum#1\endcsname{\arabic{thm}}
  \addtocounter{thm}{1}
  \expandafter\gdef\csname defkind#1\endcsname{#2}
  \expandafter\gdef\csname deflab#1\endcsname{#3}
  \expandafter\gdef\csname deflem#1\endcsname{#4}
}

\newcommand\recallfirst[1]{
  \setcounter{save}{\value{thm}}
  \setcounter{thm}{\csname deflemnum#1\endcsname}
  \begin{\csname defkind#1\endcsname}
  \label{\csname deflab#1\endcsname}
  \csname deflem#1\endcsname
  \end{\csname defkind#1\endcsname}
  \setcounter{thm}{\value{save}}
}

\newcommand\deferred[4]{
  \defer{#1}{#2}{#3}{#4}
  \recallfirst{#1}
}

\newcommand\recall[1]{
  \setcounter{save}{\value{thm}}
  \setcounter{thm}{\csname deflemnum#1\endcsname}
  \begin{\csname defkind#1\endcsname}
  \csname deflem#1\endcsname
  \end{\csname defkind#1\endcsname}
  \setcounter{thm}{\value{save}}
}

\begin{document}
\title{Bicriteria data compression}
\date{}
\author[]{Andrea~Farruggia}
\author[]{Paolo~Ferragina}
\author[]{Antonio~Frangioni}
\author[]{Rossano~Venturini}
\affil[]{Dipartimento di Informatica, Universit\`{a} di Pisa, Italy}
\affil[]{\{farruggi, ferragina, frangio, rossano\}@di.unipi.it}

\clearpage\maketitle
\thispagestyle{empty}

\begin{abstract}
	The advent of massive datasets and the consequent design of high-performing distributed storage systems---such as BigTable by Google \citep{BigTable}, Cassandra by Facebook \citep{Cassandra}, Hadoop by Apache---have reignited the interest of the scientific and engineering community towards the design of lossless data compressors which achieve effective compression ratio and very efficient decompression speed. Lempel-Ziv's \LZSE{} algorithm is the {\em de facto} choice in this scenario because its decompression is significantly faster than other approaches, and its algorithmic structure is flexible enough to trade decompression speed {\em versus} compressed-space efficiency. This algorithm has been declined in many ways, the most famous ones are: the classic {\tt gzip}, \LZF{} and Google's Snappy. Each of these implementations offers a trade-off between space occupancy and decompression speed, so software engineers have to content themselves by picking the one which comes closer to the requirements of the application in their hands.

Starting from these premises, and for the first time in the literature, we address in this paper the problem of trading optimally, and in a principled way, the consumption of these two resources by introducing and solving what we call the {\em Bicriteria \LZSE{}-Parsing problem}. The goal is to determine an \LZSE{} parsing which minimizes the space occupancy in bits of the compressed file, provided that the decompression time is bounded by a fixed amount. Symmetrically, we can exchange the role of the two resources and thus ask for minimizing the decompression time provided that the compressed space is bounded by a fixed amount. This way, the software engineer can set its space (or time) requirements and then derive the \LZSE{} parsing which optimizes the decompression speed (or the space occupancy, respectively), thus resulting the {\em best possible \LZSE{} compression under those constraints}.

We solve this problem in four stages: we turn it into a sort of {\em weight-constrained shortest path} problem (WCSPP) over a weighted graph derived from the \LZSE{}-parsing of the input file; we argue that known solutions for WSCPP are inefficient and thus unusable in practice; we prove some interesting structural properties about that graph, and then design an $O(n\log^2 n)$-time algorithm which computes a {\em small additive} approximation of the optimal \LZSE{} parsing. This additive approximation is logarithmic in the input size and thus totally negligible in practice. Finally, we sustain these arguments by performing some experiments which show that our algorithm combines the best properties of known compressors: its decompression time is close to the fastest Snappy's and \LZF{}'s, and its compression ratio is close to the more succinct {\tt bzip2}'s and {\tt LZMA}'s. Actually, in many cases our compressor improves the best known engineered solutions mentioned above, so we can safely state that with our result software engineers have an algorithmic-knob to automatically trade in a principled way the time/space requirements of their applications.

\smallskip \noindent Summarizing, the three main contributions of the paper are:
\begin{inparaenum}[(i)]
\item we introduce the novel Bicriteria \LZSE{}-Parsing problem which formalizes in a principled way what data-compressors have traditionally approached by means of heuristics;
\item we solve this problem efficiently in $O(n\log^2 n)$ time and optimal linear space, by proving and deploying some specific structural properties of the weighted graph derived from the possible \LZSE{}-parsings of the input file;
\item we execute a preliminary set of experiments which show that our novel proposal \emph{dominates} all the highly engineered competitors, hence offering a win-win situation in theory\&practice.
\end{inparaenum}
\end{abstract}

\newpage
\setcounter{page}{1}

\section{Introduction}
\label{sec:intro}

The advent of massive datasets and the consequent design of high-performing distributed storage systems---such as BigTable by Google \citep{BigTable}, Cassandra by Facebook \citep{Cassandra}, Hadoop by Apache---have reignited the interest of the scientific and engineering community towards the design of lossless data compressors which achieve effective compression ratio and very efficient decompression speed. The literature abounds of solutions for this problem, named ``compress once, decompress many times'', that can be cast into two main families: the compressors based on the Burrows-Wheeler Transform \citep{BWT94}, and the ones based on the Lempel-Ziv parsing scheme \citep{LZ77,LZ78}. Compressors are known in both families that require time linear in the input size, both for compressing and decompressing the data, and take compressed-space which can be bound in terms of the $k$-th order empirical entropy of the input \citep{LZ77,lz-entropy}.

But the compressors running behind those large-scale storage systems are not derived from those scientific results. The reason relies in the fact that theoretically efficient compressors are optimal in the RAM model, but they elicit many cache/IO misses during the decompression step. This poor behavior is most prominent in the BWT-based compressors, and it is not negligible in the LZ-based approaches. This motivated the software engineers to devise variants of Lempel-Ziv's original proposal, with the injection of several software tricks which have beneficial effects on memory-access locality. The most famous LZ-variants are the ones available for BigTable and Hadoop, namely Google's Snappy
and \LZF{}. These compressors  expanded further the known jungle of space/time trade-offs\footnote{See e.g., \url{http://cs.fit.edu/~mmahoney/compression/}}, thus posing the software engineers in front of a choice: either achieve effective/optimal compression-ratios, possibly sacrificing the decompression speed (as it occurs for the theory-based results \citep{FGMS05,FNV,OptimallyPartitioning}); or try to balance them by adopting a plethora of programming tricks which trade compressed space by decompression time (such as Snappy, \LZF{} or the recent \LZSE{}-end \citep{lzend}), thus waiving mathematical guarantees on their final performance.

In the light of this dichotomy, it would be natural to ask for an algorithm which guarantees effective compression-ratios and efficient decompression speed in hierarchical memories. In this paper, however, we aim for a more ambitious goal which is further motivated by the following two simple, yet challenging, questions:

\vspace*{-3mm}
\begin{itemize}
 \item who cares whether the compressed file is slightly longer than the one achievable with BWT-based compressors, provided that we can improve significantly BWT's decompression time? This is a natural question arising in the context of distributed storage-systems, and the one leading the design Snappy and \LZF{}.

\vspace*{-3mm}
 \item who cares whether the compressed file can be decompressed slightly slower than Snappy or \LZF{}, provided that we can improve significantly their compressed space? This is a natural question in a context where space occupancy is a major concern, e.g. tablets and smart-phones, and the one for which tools like Google's Zopfli have been recently introduced.
\end{itemize}
\vspace*{-3mm}

If we are able to offer mathematical guarantees to the meaning of ``slightly longer/slower'', then these two questions become pertinent and challenging in theory too. So in this paper we introduce the following problem, that we call \emph{bicriteria data compression}: given an input file \Txt{} and an upper bound $T$ on its decompression time, the goal is to determine a compressed version of \Txt{} which minimizes the compressed space provided that it can be decompressed in $T$ time. Symmetrically, we could exchange the role of time/space resources, and thus ask for the compressed version of \Txt{}  which minimizes the decompression time provided that the compressed-space occupancy is within a fixed bound.

In order to attack this problem in a principled way we need to fix two ingredients: the class of compressed versions of \Txt{} over which this bicriteria optimization will take place; and the computational model measuring the resources to be optimized. For the former ingredient we will take the class of {\em \LZSE{}-based compressors} because they are dominant in the theoretical (e.g., \citep{FarachT98,CohenMMSZ00,CormodeM05,KellerKLL09,FNV}) and in the practical setting (e.g., {\tt gzip}, {\tt 7zip} , Snappy, \LZF{}, \citep{lzend,Puglisi13,WittenMoffatBell99}). In Section~\ref{sec:lzse-parsing}, we will show that the Bicriteria data-compression problem formulated over \LZSE{}-based compressors is well funded because there exists an \emph{infinite class of strings} which can be parsed in many different ways, thus offering a wide spectrum of space-time trade-offs in which small variations in the usage of one resource (e.g., time) may induce arbitrary large variations in the usage of the other resource (e.g., space).

For the latter ingredient, we take inspiration from several models of computation which abstract multi-level memory hierarchies and the fetching of contiguous memory words~\citep{AggarwalHMBT,AggarwalACS87,AlpernCFS94,LuccioP93,VitterS94a}. In these models the cost of fetching a word at address $x$ takes $f(x)$ time, where $f(x)$ is a non-decreasing, polynomially bounded function (e.g., $f(x) = \lceil \log x \rceil$ and $f(x) = x^{O(1)}$). Some of these models offer also a \emph{block copy} operation, in which a sequence of $\ell$ consecutive words can be copied from memory location $x$ to memory location $y$ (with $x \geq y)$ in time $f(x) + \ell$. We remark that, in our scenario, this model is more proper than the frequently adopted two-level memory model \citep{VitterTwoLevel}, because we care to differentiate between contiguous/random accesses to memory-disk blocks, which is a feature heavily exploited in the design of modern compressors\citep{Drepper}.

Given these two ingredients, we devise a formal framework that allows us to analyze any \LZSE{}-parsing scheme in terms of both the space occupancy (in bits) of the compressed file, and the time cost of its decompression taking into account the underlying memory hierarchy (see Section~\ref{sec:lzss-parse-modeling}). More in detail, we will extend the model proposed in \citep{FNV}, based on a special weighted \texttt{DAG} consisting of $n=|\Txt{}|$ nodes, one per character of \Txt{}, and  $m = O(n^2)$ edges, one per possible phrase in the \LZSE{}-parsing of \Txt{}. In our new graph each edge will have attached two weights: a \emph{time weight}, that accounts for the time to decompress a phrase (derived according to the hierarchical-memory model mentioned above), and a \emph{space cost}, that accounts for the number of bits needed to store the \LZSE{}-phrase associated to that edge (derived according to the integer-encoder adopted in the compressor). Every path $\pi$ from node $1$ to node $n$ in \G{} (hereafter, named ``1n-path'') corresponds to an \LZSE{}-parsing of the input file \Txt{} whose {\em compressed-space occupancy} is given by the sum of the {\em space-costs} of $\pi$'s edges (say $s(\pi)$) and whose {\em decompression-time} is given by the sum of the {\em time-weights} of $\pi$'s edges (say $t(\pi)$). As a result of this correspondence, we will be able to rephrase our \emph{bicriteria \LZSE{}-parsing} problem into the well-known {\em weight-constrained shortest path problem} ({\tt WCSPP}) (see \citep{mehlhorn-rcsp} and references therein) over the weighted \texttt{DAG} \G{}, in which the goal will be to search for the 1n-path $\pi$ whose decompression-time is $t(\pi) \leq T$ and whose compressed-space occupancy $s(\pi)$ is minimized. Due to its vast range of applications, {\tt WCSPP} received a great deal of attention from the optimization community. It is an $\mathcal{NP}$-Hard problem, even on a \texttt{DAG} with positive weights and costs \citep{wcspp-nph,DynamicProg}, and it can be solved in pseudo-polynomial $O(mT)$ time via dynamic programming \citep{comb_opt}.  Our special version of the {\tt WCSPP} problem has $m$ and $T$ bounded by $O(n \log n)$ (see Section~\ref{sec:lzss-parse-modeling}), so it can be solved in polynomial time, namely  $O(mT) = O(n^2 \log^2 n)$ time and $O(n^2 \log n)$ space. Unfortunately this bounds are unacceptable in practice, because $n^2 \approx 2^{64}$ just for one Gb of data to be compressed.

The second contribution of this paper is to prove some structural properties of our weighted \texttt{DAG} which allow us to design an algorithm that approximately solves our version of {\tt WCSPP} in $O(n \log^2 n)$ time and $O(n)$ working space. The approximation is {\em additive} in that, our algorithm determines a \LZSE{}-parsing whose decompression time is $\leq T + 2\,\tmax{}$ and whose compressed space is just $\smax{}$ bits more than the optimal one, where $\tmax{}$ and $\smax{}$ are, respectively, the maximum time-weight and the maximum space-cost of any edge in the DAG. Given that the values of \smax{} and \tmax{} are logarithmic in $n$ (see Section~\ref{sec:lzse-parsing}), those additive terms are negligible. We remark here that this type of additive-approximation is clearly related to the \emph{bicriteria-approximation} introduced by \citep{marathe-biapprox}, and it is more desirable than the ``classic''  $(\alpha, \beta)$-approximation because ours is additive whereas the latter is multiplicative, so the larger is the problem size the better is our approximation. The further peculiarity of our approach is that we are using the additive-approximation to speed-up the solution to a problem that in our setting admits already a polynomial solution which, however, grows as $\Omega(n^2)$ thus resulting unusable in the practical setting.

The third, and last, contribution of this paper is to present a set of experimental results which compare an implementation of our compressor against state-of-the-art \LZSE{}-based algorithms (Snappy, \texttt{LZMA}, \LZF{}, {\tt gzip}) and  BWT-based algorithms (with bounded and unbounded memory footprint). These experiments bring out two key aspects:
\begin{inparaenum}[(i)]
\item they provide a practical ground to the two pertinent questions posed at the beginning of the paper, thus, motivating the theoretical analysis introduced with our novel Bicriteria Data-Compression problem;
\item they show that our parsing strategy dominates all the highly engineered competitors, by exhibiting decompression speeds close to those of Snappy and \LZF{} (i.e., the fastest known ones), and compression ratios close to those of BWT-based and \texttt{LZMA} compressors (i.e., the more succinct ones). This is indeed a win-win situation in theory\&practice.
\end{inparaenum}

\section{On the \LZSE{}-parsing}
\label{sec:lzse-parsing}

Let $\Txt{}$ be a string of length $n$ built over an alphabet $\Sigma=[\sigma]$, and terminated by a special character. We denote by $\Txt{}[i]$ the $i$-th character of $\Txt{}$, and by $\Txt{}[i, j]$ the substring ranging from $i$ to $j$ (included). The compression algorithm \LZSE{} works by {\em parsing} the input string $\Txt{}$ into phrases $p_1, \ldots, p_k$ such that, phrase $p_i$ can be any substring of $\Txt{}$ starting in the prefix $p_1, \ldots, p_{i-1}$. Once the parsing has been identified, each phrase is represented via \emph{codewords}, that are pairs of integers \LZP{d}{\ell}, where $d$ is the distance from the position where the copied phrase occurs, and $\ell$ is its length. Every first occurrence of a new character $c$ is encoded as \LZP{0}{c}. These pairs are compressed via variable-length integer encoders which eventually produces the compressed output of $\Txt{}$ as a sequence of bits. Among all possible parsing strategies, the \emph{greedy parsing} is widely adopted: it chooses $p_i$ as the longest prefix of the remaining suffix of $\Txt{}$. This is optimal whenever the goal is to minimize the number of generated phrases or, equivalently, the phrases have equal bit-length; but if phrases are encoded with a variable number of bits then the greedy approach may be sub-optimal \citep{FNV}.

\begin{table}[h]
\centering
\caption{Summary of main notations.}
\label{tab:defs}
\begin{tabular}{c p{.55\textwidth} p{.3\textwidth}}
\toprule
\multicolumn{1}{c}{Name}        &  \multicolumn{1}{c}{Definition}   &    \multicolumn{1}{c}{Properties}\\
\midrule
\Txt{}        &   A (null-terminated) document to be compressed.    & \\
$n$         &   Length of \Txt{} (end-of-text character included).  &   \\
$\Txt{}[i]$     &   The $i$-th character of \Txt{}. & \\
$\Txt{}[i, j]$    &   Substring of \Txt{} starting from $\Txt{}[i]$ until $\Txt{}[j]$ & \\
\LZC{c}       &   A \LZSE{} phrase which represents a single character $c$.\\
\LZP{d}{\ell}   & A \LZSE{} phrase which represents a copy of a string of length $\ell$ at distance $d$ & \\[.1cm]
$t(d)$        &   Amount of time spent in accessing the first character of a copy at distance $d$.
          &   $t(d) = O(\log n)$.\\[.2cm]
$s(d, \ell)$    &   The length in bits of the encoding of \LZP{d}{\ell}.
          &   $s(d, \ell) \leq s(d', \ell')$, for $d \leq d'$ and $\ell \leq \ell'$.\\[.2cm]
$t(d, \ell)$    &   The time needed to decompress the \LZSE{}-phrase \LZP{d}{\ell}.
          &   We have both $t(d, \ell) = t(d) + s(d, \ell)$ and \newline
            $t(d, \ell) \leq t(d', \ell')$, for $d \leq d'$ and $\ell \leq \ell'$.\\[.2cm]
$s(\pi)$      &   The space occupancy of parsing $\pi$.
          & $s(\pi) = \sum_{\LZP{d}{\ell} \in \pi} s(d, \ell)$.\\
$t(\pi)$      &   The time needed to decompress the parsing $\pi$.
          & $t(\pi) = 2n + \sum_{\LZP{d}{\ell} \in \pi} t(d, \ell)$.\\
\smax{}       & The maximum space occupancy (in bits) of any \LZSE{} phrase of \Txt{}.
          &   $\smax{} = O(\log n)$.\\
\tmax{}       &   The maximum time taken to decompress a \LZSE{} phrase of \Txt{}.
          & $\tmax{} = O(\log n)$.\\
\Scosts{}       &   The number of distinct values which may be assumed by $s(d, \ell)$ when $d \leq n, \, \ell \leq n$.
          &   $\Scosts{} = O(\log n)$.\\
\Tcosts{}     &   The number of distinct values which may be assumed by $t(d, \ell)$ when $d \leq n,\, \ell \leq n$.
          &   $\Tcosts{} = O(\log n)$.\\
\bottomrule
\end{tabular}
\end{table}

\paragraph{Modeling the space occupancy.}
A \LZSE{}-phrase \LZP{d}{\ell} is typically compressed by using two distinct (universal) integer encoders, since distances $d$ and lengths $\ell$ are distributed differently in $\Txt{}$. We use $s(d,\ell)$ to denote the length in bits of the encoding of \LZP{d}{\ell}. We restrict our attention on variable-length integer encoders which emit longer codewords for bigger integers, the so called {\em non-decreasing cost property}. This assumption is not restrictive because it encompasses all universal encoders, such as Truncated binary, Elias' Gamma and Delta~\citep{Elias}, Golomb~\citep{golomb66run}, and \LZF{}'s encoder. An interesting fact about these encoders is that they take a logarithmic number of bits per integer. This fact is crucial in evaluating the complexity of our algorithm, since it depends on the number of distinct values assumed by $s(d, \ell)$ when $d,\ell \leq n$. We denote by \Scosts{} this number, which is $O(\log n)$ for all the universal encoders above.

For the sake of presentation, we denote by $s(\pi)$ the bit-length of the compressed output generated according to the \LZSE{}-parsing $\pi$. This is estimated by summing the lengths of the encoding of all phrases in $\pi$, hence $\sum_{\langle d,\ell\rangle \in \pi} s(d,\ell)$.

\paragraph{Modeling the decompression speed.}
The aim of this section is to define a model for evaluating the time to decompress a string $\Txt{}$ compressed via \LZSE{} in a hierarchical-memory setting. The decompression proceeds from left to right in $\Txt{}$ by reconstructing one phrase at a time. For each phrase $\LZP{d}{\ell}$, the decompressor needs to decode its codeword and then copy the substring of length $\ell$ at distance $d$ from the current position in $\Txt{}$. In terms of memory accesses this means a random access to locate that copy plus the cost of reading it. Taking inspiration from models in \citep{AggarwalHMBT,AggarwalACS87,AlpernCFS94,LuccioP93,VitterS94a}, we assume that accessing a character at distance $d$ takes $t(d)$ time, where $t(d)=\lceil \log d \rceil$,  whereas scanning $\ell$ consecutive characters takes $\ell$ time regardless of the memory level containing these characters.\footnote{See Drepper's monograph on memory hierarchies \citep{Drepper}.}

Under these assumptions, the decompression of a phrase $\LZP{d}{\ell}$ takes $s(d,\ell)$ time to read and decode the codeword of $d$ and $\ell$, time $t(d)+\ell$ to read the copy, and time $\ell$ to append it to $\Txt{}$. Summing over all phrases we get a total decompression time of $t(\pi)=2 n + \sum_{\LZP{d}{\ell} \in \pi} (t(d) + s(d,\ell))$. Since the $2n$ term is independent of the parsing it can be neglected, thus focussing on the terms $t(d, \ell)=t(d) + s(d,\ell)$ for each individual phrase of $\pi$. As in the previous section we denote by \Tcosts{} the number of distinct values which may be assumed by $t(d, \ell)$ when $d,\ell \leq n$; clearly $\Tcosts{} = O(\log n)$. Similarly to \Scosts{}, this term will be crucial in defining the time complexity of our algorithm.

\paragraph{Pathological strings: space/time trade-offs matter.}
In our context we are interested in \LZSE{}-parsings which ``optimize'' two criteria, namely \emph{decompression time} and \emph{compressed space}. In this respect, the notion of ``best'' parsing needs to recall the one of \emph{Pareto-optimal} parsings, i.e., parsings which are not worse than some others in one parameter, being it the decompression time or the compressed space. The key new result here is to show, as claimed in the introduction, that there exists an infinite family of strings for which the Pareto-optimal parsings exhibit significant differences in their decompression time versus compressed space.

For the sake of presentation let us assume that each codeword takes constant space, and that our model of computation consists of just two memory levels such that the access time of the fastest level (of size $c$) is negligible, while the access time of the slowest level (of unbounded size) is substantial. We construct our pathological input string $\Txt$ as follows. Fix any string $P$ of length at most $c$ drawn over a alphabet $\Sigma$ which can be \LZSE{}-parsed with $k$ phrases. For any $i \geq 0$, let $B_i$ be the string $\$^{c+i}P$ with $\$$ a special symbol not in $\Sigma$.
Our string $\Txt$ is $B_0B_1\ldots B_m$. Since the length of run of $\$$s increases as $i$ increases, no pair of consecutive strings $B_{i}$ and $B_{i+1}$ can be part of the same \LZSE{}-phrase. Moreover, we have two alternatives in parsing each $B_i$, with $i \geq 1$: (1) we parse $B_i$ by deploying only its content and thus not requiring any cache miss at decompression time, this uses $2+k$ phrases which copy at distance at most $c$; (2) we parse $B_i$ by using $2$ phrases copied from the previous string $B_{i-1}$, thus requiring one cache miss at decompression time.

There are $m-1$ Pareto-optimal parsings of $\Txt$  obtained by choosing one of the above alternatives for each string $B_i$. On one extreme, the parser always chooses alternative (1) obtaining a parsing with $m(2+k)$ phrases which is decompressible with no cache misses. On the other extreme, the parser always prefers alternative (2) obtaining a parsing with $2+k+2m$ phrases which is decompressible with $m-1$ cache misses. In between these two extremes, we have a plethora of Pareto-optimal parsings: we can move from one extreme to the other by trading decompression speed for space occupancy. In particular, we can save $k$ phrases at the cost of one more cache miss, where $k$ is a value which can be varied by choosing different strings $P$. The ambitious goal of this paper is to automatically and efficiently choose any of these trade-offs.

\section{From \LZSE{}-Parsing to a weighted DAG}
\label{sec:lzss-parse-modeling}

In this section we model the \emph{bicriteria \LZSE{}-parsing} problem as a \emph{Weight-Constrained Shortest Path} problem ({\tt WCSPP}) over a weighted DAG \G{} defined as follows. Given an input string $\Txt{}$ of length\footnote{Recall that \Txt\ is terminated by a special character.} $n$,the graph \G{} consists of $n$ nodes, one per input character, and $m$ edges, one per possible \LZSE{}-phrase in \Txt{}. In particular we distinguish two types of edges: $(i,i+1)$, which represents the case of the single-character phrase $\langle 0,\Txt{}[i]\rangle$, and $(i,j)$ with $j = i+\ell>i+1$, which represents the phrase $\LZP{d}{\ell}$ and thus the case of $\Txt{}[i,i+\ell-1]$ occurring $d$ characters before in \Txt{}. This construction was proposed in \citep{LZOptimal}: clearly, \G{} is a DAG and each path from node $1$ to node $n$ (1n-path) corresponds to an \LZSE{}-parsing of \Txt{}. Subsequently, \citep{FNV} added the weight $s(d,\ell)$ to the edge $(i,j)$ in order to denote its space occupancy in bits.

We extend this modeling by adding another weight to \G{}'s edges, namely the time $t(i,j)$ taken to decode $\LZP{d}{\ell}$. This way, every 1n-path $\pi$ not only identifies an \LZSE{}-parsing of \Txt{}, but also the sum of the space-costs ($s(\pi)$) and the sum of the time-weights ($t(\pi)$) of its edges define its compressed bit-space occupancy and its decompression time, respectively. As a result of this modeling, we can re-phrase our bicriteria \LZSE{}-parsing as the Weighted-Constrained Shorted Path problem in \G{}, which asks for $\min_{\pi \in \Pi} \; s(\pi)$ provided that $t(\pi) \leq T$. Clearly we could reverse the role of space and time in \G{}'s edges, but for ease of explanation, in the rest of the paper we will consider only the first formulation, even if our algorithmic solution can be used for both versions without any loss in its time/space efficiency.

\medskip
In the following, we say that an edge $(i',j')$ is \emph{nested} in an edge $(i,j)$ whenever $i \leq i' < j' \leq j$. To design efficient algorithms for {\tt WCSPP}, it is crucial to exploit the peculiar properties of \G{}.
\begin{Property}
\label{prt}
Given an edge $(i,j)$ of \G{}, any $(i',j')$ nested in $(i,j)$ is
\begin{inparaenum}[(\itshape a)]
\item an edge of \G{} and
\item its time- and space-weights are smaller or equal than the ones of $(i,j)$.
\end{inparaenum}
\end{Property}

The first property derives from the fact that \G{} models the parsing of a text using a prefix-/suffix-complete dictionary, as the \LZSE{} one. The second property derives from the fact that the functions $s(d,\ell)$ and $t(d,\ell)$, which model the time/space edge-weights, are non-decreasing in both arguments. So, given a phrase $\Txt{}[i,j]$ and its corresponding codeword \LZP{d}{\ell}, any substring $\Txt{}[i,j']$ is also a phrase (from the prefix-complete property) and its codeword $\LZP{d'}{\ell'}$ is such that $d' \leq d$ and $\ell' \leq \ell$, because  $\Txt{}[i,j']$ occurs at least wherever $\Txt{}[i,j]$ does.

\subsection{Pruning the graph}
\label{sec:pruning}
The size of \G{} may be quadratic in $n$; just consider the string $\Txt{} = a^n$ which generates one edge per substring of \Txt{}. Given that $n$ is typically of the order of millions or even billions, storing the whole \G{} is unfeasible. This problem has been already faced in \citep{FNV} while solving the bit-optimal \LZSE{}-parsing problem over a graph with only the space-cost edges. Their solution mainly relied on two ideas: (i) pruning from their graph a large subset of unnecessary edges, yet guaranteeing that the bit-optimal path is preserved, and (ii) generating the forward stars of the nodes in the pruned graph on-the-fly by means of an algorithm, called \texttt{FSG}. It was shown in \citep{FNV} that such pruned graph has size $O(n \, \Scosts{})$ and can be generated incrementally in that time and only $O(n)$ space.

The key contribution of this section is twofold: we show that there exists a small subgraph of \G{}, consisting of $O(n (\Scosts{} + \Tcosts{}))$ edges, which includes all Pareto-optimal 1n-paths of \G{}; we then show that this pruned graph can be generated efficiently by using the \texttt{FSG} algorithm. The monotonicity property stated in Property~\ref{prt} for the $s()$-cost and the $t()$-weight of DAG-edges allows us to define the notion of \emph{maximality} of an edge, which (in turn) is correlated to the property of Pareto-optimality of a 1n-path in \G{}.

\begin{Definition}
An edge $e=(i,j)$ is said to be \emph{$s$-maximal} iff, either the (next) edge $e'=(i,j+1)$ does not exist, or it does exist but the $s$-cost of $e'$ is strictly larger than the $s$-cost of $e$. In a similar vein we define the notion of \emph{$t$-maximal} edge, and state that an edge is \emph{maximal} whenever it is either \emph{$s$-maximal} or \emph{$t$-maximal}, or both.
\end{Definition}

Lemma~\ref{lmm:path-suffix} shows that, for any path $\pi$ from node $i$ to $n$ and for each $i' > i$, there is a path from $i'$ to $n$ with cost/time not higher than those of $\pi$. 

\deferred{path-suffix}{Lemma}{lmm:path-suffix}{
For each triple of nodes $i < i' < j$, and for each path $\pi$ from $i$ to $j$, there exists a path $\pi'$ from $i'$ to $j$ such that $t(\pi') \leq t(\pi)$ and $s(\pi') \leq t(\pi)$.
}
\begin{proof}
Let $(h,k)$ be the edge of $\pi$ which surpasses $i'$ in \G{}, i.e, $h < i' \leq k$, and let $\pi''$ be the sub-path of $\pi'$ from $k$ to $j$. If $i' = k$, the thesis follows by setting $\pi' = \pi''$, and noticing that this is a suffix subpath of $\pi$ thus incurring in smaller costs. Otherwise, the edge $(i',k)$ exists (because of the suffix-completeness property of \LZSE{}-phrases), and its time and space weights are not greater than the corresponding ones of edge $(h,k)$ (Property~\ref{prt}). Thus the thesis follows by setting $\pi' = (i',k) \cdot \pi''$.
\end{proof}

The lemma stated above allows to ``push'' to the right non-maximal edges by iteratively substituting non-maximal edges with maximal ones without augmenting the time and space costs of the path. This fact is exploited in Theorem~\ref{thm:pruning}, which shows that the search of optimal paths in \G{} can be limited to those composed of maximal edges only.

\deferred{pruning}{Theorem}{thm:pruning}{
For any 1n-path $\pi$ there exists a 1n-path $\pi^\star$ composed of maximal edges only and such that $\pi^\star$ is not worse than $\pi$ in any one of its two costs, i.e., $t(\pi^\star) \leq t(\pi)$ and $s(\pi^\star) \leq s(\pi)$.
}
\begin{proof}
We show that any 1n-path $\pi$ containing non-maximal edges can be turned into a 1n-path $\pi'$ containing maximal edges only. Take the leftmost non-maximal edge in $\pi$, say $(v,w)$, and denote by $\pi_v$ and $\pi_w$, respectively, the prefix/suffix of path $\pi$ ending in $v$ and starting from $w$. By definition of maximality, it must exist a maximal edge $(v,z)$, with $z > w$, whose time/space weights are the same ones of $(v,w)$. We can then apply Lemma~\ref{lmm:path-suffix} to the triple $(w,z,n)$ and thus derive a path $\mu$ from $z$ to $n$ such that $s(\mu) \leq s(\pi_w)$ and $t(\mu) \leq t(\pi_w)$.

We then construct the 1n-path $\pi''$ by connecting the sub-path $\pi_v$, the maximal edge $(v,z)$, and the path $\mu$: using Lemma~\ref{lmm:path-suffix} one readily shows that the time/space costs of $\pi''$ are not larger than these of $\pi$. The key property is that we pushed right the leftmost non-maximal edge (if any), which must now occur (if ever) within $\mu$; by iterating this argument we get the thesis.
\end{proof}

Let $\widetilde{\G{}}$ be the pruned graph defined by keeping only maximal edges in \G{}. Since the set of maximal edges is given by the union of $s$-maximal and $t$-maximal edges, there cannot be more than $\Scosts{} + \Tcosts{}$ maximal edges outgoing from any node. Given that both \Scosts{} and \Tcosts{} are $O(\log n)$, it follows that $\widetilde{\G{}}$ has at most $O(n \log n)$ edges, and thus it is asymptotically sparser than \G{}. Due to lack of space we cannot dig into the generation of these edges (details in the journal paper), so here we state that all maximal edges of \G{} can be generated on-the-fly by easily adapting the \texttt{FSG}-algorithm \citep{FNV} and taking $O(1)$ amortized time per edge, hence overall $O(n \log n)$ time and $O(n)$ bits of working space. This is surprising because it means that the retrieval of the optimal path $\pi^\star$ can be done by examining only a (significantly smaller) sub-graph of \G{} which can be generated in an optimal output-sensitive manner.

\section{Our Approximation Algorithm}
\label{sec:algorithm}

This section is devoted to solve {\tt WSCCP} over the weighted DAG \G{} whose structure and weights satisfy Property~\ref{prt}. Recall that $\tmax{}$ and $\smax{}$ are, respectively, the maximum time-cost and the maximum space-weight of the edges in \G{}. We denote with $z(P)$ the optimal value of an optimization problem $P$, set $\varphi^\star = z({\tt WCSPP})$, and use ${\tt WCSPP}(\lambda)$ to denote the Lagrangian relaxation of {\tt WCSPP} with Lagrangian multiplier $\lambda$, namely:

\begin{displaymath}
	\tag{\ensuremath{{\tt WCSPP}\left( \lambda \right)}} \min_{\pi \in \Pi} \; s(\pi) + \lambda(t(\pi) - T).
\end{displaymath}

As mentioned in the introduction, our algorithm works in two phases. In the first phase, described in Section~\ref{sec:dual}, the algorithm solves the Lagrangian Dual problem through a specialization of Kelley's cutting-plane algorithm \citep{kelley-cp}, as first introduced by Handler and Zang \citep{zang-dual}. The result is a lower-bound $z^\star$ for {\tt WCSPP} and an instantiation for the parameter $\lambda^\star \geq 0$ which maximizes the optimal value of ${\tt WCSPP}(\lambda)$. In addition, this computes a pair of paths $(\pi_L, \pi_R)$ which are optimal for ${\tt WCSPP}(\lambda^\star)$ and are such that $t(\pi_L) \geq T$ and $t(\pi_R) \leq T$.

In case one path among them satisfies the time bound $T$ exactly, then its space-cost equals the optimal value $\varphi^\star$, and thus that path is an optimal solution for \texttt{WSCPP}. Otherwise, the algorithm starts the second phase, described in Section~\ref{sec:path-swap}, which is the more technical algorithmic contribution of this paper. This phase derives a new path by joining a proper prefix of $\pi_L$ with a proper suffix of $\pi_R$. The key difficulty here is to show that this new path guarantees an additive-approximation of the optimal solution, and it can be computed in just $O(n)$ time. At the end, we will have proved the following theorem.

\begin{Theorem}
\label{thm:main}
There is an algorithm which computes a path $\pi$ such that $s(\pi) \leq \varphi^\star + \smax{}$ and $t(\pi) \leq T + 2\,\tmax{}$ in $O(n\; \log n \; \log(n \; \tmax{} \; \smax{}))$ time and $O(n)$ space.
\end{Theorem}

We call this type of result an $(\smax{}, 2\,\tmax{})$-additive approximation. By recalling that $\smax{}$ and $\tmax{}$ are $O(\log n)$, since we are using universal integer encoders and memory hierarchies whose time access grows logarithmically (see Section~\ref{sec:lzse-parsing}), it holds:

\begin{Corollary}
\label{cor:main}
There is an algorithm that computes an $(O(\log n), O(\log n))$-additive approximation of the Bicriteria data-compression problem in $O(n\; \log^2 n)$ time and $O(n)$ space.
\end{Corollary}

It is important to remark that this type of approximation is very strong because it is {\em additive} rather than {\em multiplicative} in the value of the bounded resources, as instead occur for the  ``classic''  $(\alpha, \beta)$-approximation \citep{handbook-apx}. This additive-approximation \emph{improves} as the value of the optimal solution grows, conversely to what occurs in the multiplicative-approximation for which, as the optimum grows, the error grows too. Actually, very few additive approximation algorithms (otherwise known as \emph{absolute approximation algorithms}) are known \citep{handbook-apx}, since many $\mathcal{NP}$-Hard problems do not admit such algorithms unless $\mathcal{NP} = \mathcal{P}$. Therefore our result gains significance from this complexity-theory perspective, too.

Interestingly, from Theorem \ref{thm:main} we can derive a FPTAS for our problem as stated in the following theorem and proved in Appendix~\ref{sec:multiplicative-apx}.

\deferred{multiplicative-apx}{Theorem}{thm:multiplicative-apx}{
For any fixed $\epsilon > 0$, then there exists a multiplicative $\left(\epsilon, \frac{\epsilon}{2}\right)$-approximation scheme for {\tt WCSPP} which takes $O\left(\frac{1}{\epsilon} \left(n \; \log^2 n + \frac{1}{\epsilon^2} \log^4 n\right) \right)$ time and $O(n + \frac{1}{\epsilon^3} \log^4 n)$ space complexity.
}

\noindent Notice that by setting $\epsilon > \sqrt[3]{\frac{\log^4 n}{n}}$, the bounds become $O\left(\frac{1}{\epsilon} n \; \log^2 n \right)$ time and $O(n)$ space.

\subsection{First phase: The cutting-plane algorithm}
\label{sec:dual}
The first phase consists of solving the Lagrangian dual of problem \texttt{WCSPP} through the first phase of Handler and Zang's seminal paper \citep{zang-dual}. Our key observation is that each iteration can be implemented by solving a bit-optimal \LZSE{}-problem formulated over the pruned graph $\widetilde{\G{}}$.

The Lagrangian dual of problem {\tt WCSPP} is $\max_{\lambda \geq 0}\;\min_{\pi \in \Pi}\; s(\pi) + \lambda (t(\pi) - T)$. This can be rewritten as a (very large) linear program in which every 1n-path defines one of the constraints and, possibly, one face of the feasible region:  $\max_{\lambda \geq 0} \left\{\,u\,:\, u \leq s(\pi) + \lambda(t(\pi) - T),\;\forall \pi \in \Pi\,\right\}$.

This can be interpreted geometrically. Let us denote as $L(\pi, \lambda)$, or \emph{$\lambda$-cost}, the Lagrangian cost $s(\pi) + \lambda(t(\pi) - T)$ of the path $\pi$ with parameter $\lambda$. Each path $\pi$ represents thus the line $\varphi = L(\pi, \lambda)$ in the Euclidian space $(\lambda, \varphi)$.
Feasible paths have a non-positive slope (since $t(\pi) \leq T)$, unfeasible paths have a positive slope (since $t(\pi) > T$). Let us now consider the Lagrangian function $\varphi(\lambda) = \min_{\pi \in \Pi} L(\pi, \lambda)$. This function is piecewise linear and represents the lower envelope of all the “lines” in $\Pi$. A convenient way of interpreting the large linear program above is as the problem of maximizing the function $\varphi(\lambda)$ over all $\lambda \geq 0$. Unfortunately, the exponential number of paths makes impossible to solve this by a brute-force approach. However, the full set of paths $\Pi$ is not needed. In fact, we can use a \emph{cutting-plane} method \citep{kelley-cp} which determines
a pair of paths $(\pi_L, \pi_R)$ such that
\begin{inparaenum}[(i)]
    \item $L(\pi_L, \lambda^\star) = L(\pi_R, \lambda^\star) = $ the optimal (maximum) value of $\varphi(\lambda)$ and
    \item $t(\pi_L) \geq T$ and $t(\pi_R) \leq T$
\end{inparaenum}.
Referring to the terminology often used to describe the simplex method \citep{nelder1965simplex}, these paths correspond to a (feasible) \emph{optimal basis} of the linear program. 

\begin{figure}
    \centering
    \tikzstyle{pline}=[very thin, color=gray]
	\tikzstyle{label}=[font=\Large]
	\tikzstyle{slabel}=[font=\small]
	\begin{tikzpicture}
		\draw[pline] (0,0.2) -- (3,4);
		\draw[pline] (0,1) -- (5,4);
		\draw[pline] (0,2) node[slabel, color=black, anchor=south west] {$\pi_1$} -- (8,3);
		\draw[pline] (0,3.5) node[slabel, color=black, anchor=south west] {$\pi_2$} -- (8,2);
		\draw[pline] (4,4) -- (8,1);

		\draw[thick,color=red] (0,0.2) --
								(intersection of 0,0.2--3,4 and 0,1--5,4) --
								(intersection of 0,1--5,4 and 0,2--8,3) -- node[slabel,anchor=north,color=black] {$\varphi_B(\lambda)$}
								(intersection of 0,2--8,3 and 0,3.5--8,2) --
								(intersection of 0,3.5--8,2) and 4,4--8,1) --
								(intersection of 4,4--8,1 and 8,0--8,4);	
		\draw[->,thick] (0,0) -- node[label, anchor=east] {$\varphi$} (0,4) ;
		\draw[->,thick] (0,0) -- node[label, anchor=north] {$\lambda$} (8,0);

		\draw[dotted] (intersection of 0,2--8,3 and 0,3.5--8,2)-- (intersection of 0,2--8,3 and 0,3.5--8,2) |- (0,0);
		\node (pt) at (intersection of 0,2--8,3 and 0,3.5--8,2) {};
		\node[slabel, anchor=north] at (intersection of pt--[yshift=-10cm]pt and 0,0--6,0) {$\lambda^+$};
	\end{tikzpicture}

    \caption{Each path $\pi \in B$ is a line $\varphi = L(\pi, \lambda)$, and $\varphi_B(\lambda)$ (in red) is given by the lower envelope of all the lines in the space.}
    \label{fig:phi-lambda1}
\end{figure}
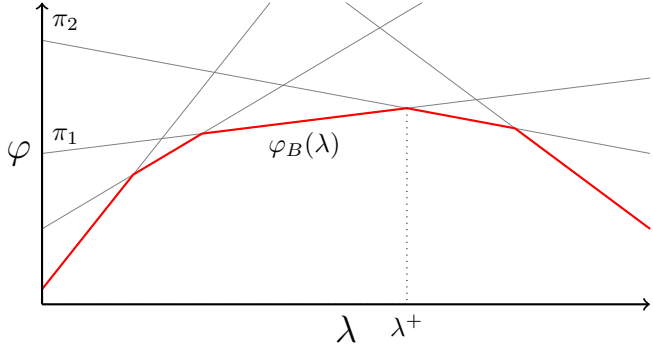

The cutting-plane method is iterative and best explained geometrically. At each step, the algorithm keeps a set $B$ of 1n-paths. At the beginning $B$ is given by the \emph{space-optimal} and the \emph{time-optimal} paths, which can be obtained by means of two shortest path computations over $\widetilde{\G{}}$. Set $B$ defines the restricted Lagrangian function $\varphi_B(\lambda)$ which is a restriction of the function $\varphi(\lambda)$ to the paths $B \subseteq \Pi$, as illustrated in Figure~\ref{fig:phi-lambda1}. The maximum value of the function $\varphi_B(\lambda)$ is identified by two paths $\pi_1, \pi_2$ in $B$, with $\pi_1$ having a non-negative slope (thus $t(\pi_1) \geq T$) and $\pi_2$ having a non-positive slope (hence $t(\pi_2) \leq T$). Let $\lambda^+$ be the intersection point between $\pi_1$ and $\pi_2$, as illustrated in Figure~\ref{fig:phi-lambda1}. Since $\varphi(\lambda)$ may be interpreted as the lower envelope of a set of lines given by paths in $\Pi \supseteq B$, it holds $\varphi_B(\lambda) \geq \varphi(\lambda)$ for each $\lambda \geq 0$. As a corollary $\varphi_B(\lambda^+) \geq \varphi(\lambda^\star)$, i.e., the optimal value of $\varphi_B(\lambda)$ is an upper-bound to the optimal value of $\varphi(\lambda)$. In particular, $\varphi_B(\lambda^+)$ is strictly greater than $\varphi(\lambda^\star)$ when $B$ does not contain an optimal basis.

At each step, the algorithm knows the value $\lambda^+$ (by induction) which maximizes $\varphi_B \left(\lambda \right)$, for the current subset $B$. Then it computes a path $\pi^+$ for which $L(\pi^+, \lambda^+) = \varphi\left( \lambda^+ \right) = \min_{\pi \in \Pi} L(\pi, \lambda^+)$ (according to definition of $\varphi$). Our key observation here is that path $\pi^+$ can be determined by searching for a shortest path whose (relaxed) cost is evaluated as $s(\pi) + \lambda^+ t(\pi)$ within the pruned DAG $\widetilde{\G{}}$. Nicely, this search can be implemented via an adaptation of the \texttt{FSG}-algorithm \citep{FNV} thus taking $O(n \log n)$ time and $O(n)$ space (see above).

In the case that the computed $\varphi \left( \lambda^+ \right) = L(\pi^+, \lambda^+)$ equals $\varphi_B \left( \lambda^+ \right)$ (which is known by induction) then the pair $(\pi_1, \pi_2)$ is an optimal basis, and the algorithm stops by setting $\lambda^\star = \lambda^+$ and $(\pi_L, \pi_R) = (\pi_1, \pi_2)$. Otherwise, the algorithm adds $\pi^+$ to $B$ and it maintains the induction by setting $(\pi_1$, $\pi_2)$ and $\lambda^+$ to reflect the new optimal value of $\varphi_B$. A simple geometric argument shows that $(\pi_1, \pi_2)$ can be computed as $(\pi, \pi_R)$, if $\pi$ is unfeasible, or by $(\pi_L, \pi)$ otherwise.

The last question is for how many iterations we have to run the cutting-plane algorithm above. Mehlhorn and Ziegelmann have shown \citep{mehlhorn-rcsp} that, for the case where the costs and the resources of each arc are integers belonging to the compact sets $[0, C]$ and $[0, R]$ respectively, then the cutting-plane algorithm (which they refer to as the \emph{Hull approach}) terminates in $O(\log(nRC))$ iterations. In our context $R=C=O(n)$:

\begin{Lemma}
\label{lem:cp-complexity}
The first phase computes a lower-bound $z^\star$ for {\tt WCSPP}, an instantiation for $\lambda^\star \geq 0$ which maximizes the optimal value of ${\tt WCSPP}(\lambda)$, and a pair of paths $(\pi_L, \pi_R)$ which are optimal for ${\tt WCSPP}(\lambda^\star)$. This takes $O(\tilde{m} \log(n \; \tmax{} \; \smax{}))$ time and $O(n)$ space, where $\tilde{m}=O(n \log n)$ is $\widetilde{\G{}}$'s size.
\end{Lemma}

\subsection{Second phase: The path-swapping algorithm}
\label{sec:path-swap}

We notice that the solution computed with Lemma \ref{lem:cp-complexity} cannot be bounded in terms of the space-optimal solution of WCSPP. Therefore the second phase of our algorithm is the technical milestone that allows to turn the basis $(\pi_L, \pi_R)$ into a path whose time- and space-costs can be mathematically bounded in terms of the optimal solution for WCSPP. In the following we denote a path as a sequence of increasing node-IDs and do not allow a node to appear multiple times in a path, so a path $(v, w, w, w, z)$ must be intended as $(v, w, z)$. Moreover, we use the following notation.
\begin{itemize}
 \item $\Pre(\pi,v)$ is the {\em prefix} of a 1n-path $\pi$ ending into the largest node $v' \leq v$ in $\pi$.
 \item $\Su(\pi,v)$ is the \emph{suffix} of a 1n-path $\pi$ starting from the smallest node $v'' \geq v$ in $\pi$.
\end{itemize}
Given two paths $\pi_1$ and $\pi_2$ in \G{}, we call {\em path swapping} through a \emph{swapping-point} $v$, which belongs either to $\pi_1$ or $\pi_2$ (or both), the operation which creates a new path, denoted by $\paths{\pi_1}{\pi_2}{v} = (\Pre(\pi_1,v), v, \Su(\pi_2,v) )$, that connects a prefix of $\pi_1$ with a suffix of $\pi_2$ via $v$.

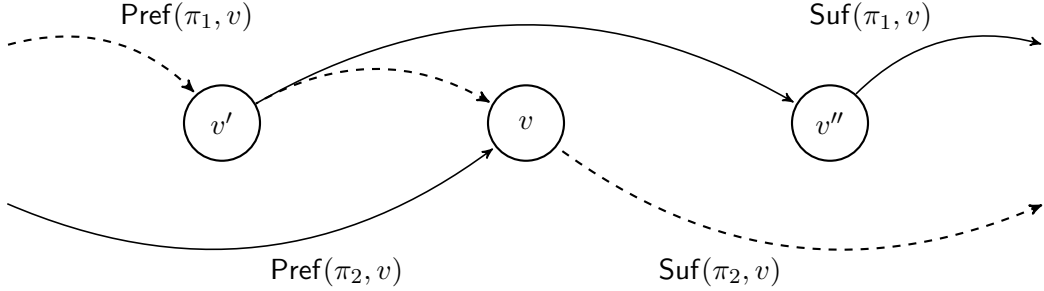
\begin{figure}
    \centering
	\tikzstyle{vertex}=[circle,draw=black,thick, minimum size=10mm]
	\tikzstyle{pre}=[<-,shorten <=1pt,>=stealth',semithick]
	\tikzstyle{post}=[->,shorten >=1pt,>=stealth',semithick]
	\begin{tikzpicture}[node distance=4cm]
		\node	[vertex] (v')					{$v'$};
		\node	[circle] (vslu)	at ([xshift=-3cm, yshift=1cm] v') {}
			edge [post, dashed, thick, bend left]	node[auto] {$\Pre(\pi_1, v)$}(v');
		\node	[circle] (vslb)	at ([xshift=-3cm, yshift=-1cm] v') {};
		\node	[vertex] (v)	[right of=v']	{$v$}
			edge [pre, dashed, thick, bend right]	(v')
			edge [pre,bend left] 	node[auto] {$\Pre(\pi_2, v)$}(vslb);
		\node	[vertex] (v'')	[right of=v] 	{$v''$}
			edge [pre, bend right]	(v');
		\node	[circle] (vsru)	at ([xshift=3cm, yshift=1cm] v'') {}
			edge [pre, bend right] 	node[auto,anchor=south east] {$\Su(\pi_1, v)$}(v'');
		\node	[circle] (vsrb)	at ([xshift=3cm, yshift=-1cm] v'') {}
			edge [pre, dashed, thick, bend left] node[auto] {$\Su(\pi_2, v)$}(v);
	\end{tikzpicture}
    \caption{A path-swap of $\pi_1$, $\pi_2$ at the swapping point $v$. The resulting path is dashed. }
    \label{fig:path-swap}
\end{figure}

Property~\ref{prt} guarantees that the path-swap operation is well-defined and, in fact, the next Fact~\ref{fact:well-defined} states that we always have edges to connect the last node of $\Pre(\pi_1,v)$ with $v$, and $v$ with the first node of $\Su(\pi_2,v)$. An illustrative example is provided in Figure~\ref{fig:path-swap}.

\begin{Fact}
\label{fact:well-defined}
    The path-swap operation is well-defined for each pair of 1n-paths $(\pi_1, \pi_2)$ and for each swapping-point $v$, which belongs either to $\pi_1$ or $\pi_2$ (or both).
\end{Fact}


For any given $\lambda \geq 0$, a path $\pi$ is \emph{$\lambda$-optimal} if its Lagrangian cost $L(\pi, \lambda)$ is equal to the value of the Lagrangian function $\varphi(\lambda)$. The following lemma shows that any path-swap of two $\lambda$-optimal paths is off at most \tmax{} in time and \smax{} in space from being a $\lambda$-optimal path.

\deferred{paths-lagrange}{Lemma}{lmm:paths-lagrange}{
  Let $\pi_1$, $\pi_2$ be $\lambda$-optimal paths, for some $\lambda \geq 0$.
  Consider the path $\pi_A = \paths{\pi_1}{\pi_2}{v}$, where $v$ is an arbitrary swapping point.
  There exist values $s, t$ such that $s \leq s(\pi_A) \leq s + \smax{}$, $t \leq t(\pi_A) \leq t + \tmax{}$ and $s + \lambda(t - T) = \varphi(\lambda)$.
}
\begin{proof}
Let $\pi_B = \paths{\pi_2}{\pi_1}{v}$: we claim that
\[
 L(\pi_A, \lambda) + L(\pi_B, \lambda) \leq
 2 \varphi(\lambda) + \smax{} + \lambda\,\tmax{}
 \;\; ,
\]
which then immediately gives the thesis since $\varphi(\lambda) \leq L(\pi, \lambda)$ for each 1n-path $\pi$.

Let us denote $\ell(i, j)$ as the scalarized cost $s(i, j) + \lambda,t(i,j)$ of edge $(i,j)$, and use $\ell(\pi)=\sum_{(i, j) \in \pi} \ell(i, j)$ as the sum of the scalarized costs of all edges in $\pi$, so that $L(\pi, \lambda) = \ell(\pi) - \lambda T$. Moreover, let us use the notation $P_j = \ell(\Pre(\pi_j, v))$ and $S_j = \ell(\Su(\pi_j, v))$ for, respectively, the scalarized costs of the prefix and suffix of the path $\pi_j$ before/after the swapping point $v$. There are three cases to consider:
\begin{enumerate}
 \item \emph{$v$ belongs to both $\pi_1$ and $\pi_2$}: In this case, we have $\ell(\pi_A) = P_1 + S_2$, $\ell(\pi_B) = P_2 + S_1$, $\ell(\pi_1) = P_1 + S_1$ and $\ell(\pi_2) = P_2 + S_2$. Since $\ell(\pi_1) + \ell(\pi_2) = \ell(\pi_A) + \ell(\pi_B)$ and $\pi_1$ and $\pi_2$ are $\lambda$-optimal paths, we have $L(\pi_A, \lambda) + L(\pi_B, \lambda) = L(\pi_1, \lambda) + L(\pi_2, \lambda) = 2 \varphi(\lambda)$ from which our claim follows (with equality).
 \item \emph{$v$ does not belong to $\pi_1$}: let $v'$ and $v''$ be, respectively, the rightmost node preceding $v$ and the leftmost node following $v$ in $\pi_1$ (see Figure~\ref{fig:path-swap}).
       %
       %
       We have
       \begin{itemize}
        \item $\ell(\pi_1) = P_1 + \ell(v', v'') + S_1$;
        \item $\ell(\pi_2) = P_2 + S_2$;
        \item $\ell(\pi_A) = P_1 + \ell(v', v) + S_2$;
        \item $\ell(\pi_B) = P_2 + \ell(v, v'') + S_1$.
       \end{itemize}
       By using the above relations we have
       \[
        \ell(\pi_A) + \ell(\pi_B) =
        P_1 + \ell(v', v) + S_2 + P_2 + \ell(v, v'') + S_1 =
        \ell(\pi_1) + \ell(\pi_2) -
        \ell(v', v'') + \ell(v', v) + \ell(v, v'')
       \]
       which then gives our claim observing that $\pi_1$ and $\pi_2$ are $\lambda$-optimal paths, $\ell(v', v) \leq \ell(v', v'')$ due to the non-decreasing cost property, and $\ell(v, v'') \leq \smax{} + \lambda \tmax{}$.
 \item \emph{$v$ does not belong to $\pi_2$}: this case is symmetric to the previous one.
\end{enumerate}
\end{proof}

Now, consider two paths $\pi_1$, $\pi_2$ to be swapped and two \emph{consecutive swapping points}, that is, two nodes $v$ and $w$ belonging to either $\pi_1$ or $\pi_2$ and such that there is no node $z$ belonging to $\pi_1$ or $\pi_2$ with $v < z < w$. 
The lemma below states that time and space of paths \paths{\pi_1}{\pi_2}{v} and \paths{\pi_1}{\pi_2}{w} differ by at most \tmax{} and \smax{}.

\deferred{paths-step}{Lemma}{lmm:paths-step}{
 Let $\pi_1$, $\pi_2$ be two paths to be swapped. Let also $v$ and $w$ be two consecutive swapping points. Set $\pi = \paths{\pi_1}{\pi_2}{v}$ and $\pi' = \paths{\pi_1}{\pi_2}{w}$: then, $|s(\pi) - s(\pi')| \leq \smax{}$ and $|t(\pi) - t(\pi')| \leq \tmax{}$.
}
\begin{proof}
 Let us consider the sub-paths $\Pre = \Pre(\pi, v)$ and $\Pre' = \Pre(\pi', w)$. There are two cases:
 \begin{enumerate}
  \item $v \in \pi_1$: in this case, $\Pre' = (\Pre, w)$. Thus, $s(\Pre') - s(\Pre) = s(v, w)$ and $t(\Pre') - t(\Pre) = t(v, w)$;
  \item $v \notin \pi_1$: let $\Pre = (v_1, \ldots, v_k, v)$; in this case, we have $\Pre' = (v_1, \ldots, v_k, w)$. Thus, we have $s(\Pre') - s(\Pre) = s(v_k, w) - s(v_k, v) \leq \smax{}$; a similar argument holds for the time weight.
 \end{enumerate}
 Thus, $s(\Pre') - s(\Pre) \leq \smax{}$ and $t(\Pre') - t(\Pre) \leq \tmax{}$. Symmetrically, it holds $s(\Su) - s(\Su') \leq \smax{}$ and $t(\Su) - t(\Su') \leq \tmax{}$; since $s(\pi) = s(\Pre) + s(\Su)$ and $s(\pi') = s(\Pre') + s(\Su')$, it follows $|s(\pi) - s(\pi')| \leq \smax{}$ , and a similar argument holds for $|t(\pi) - t(\pi')|$.
\end{proof}

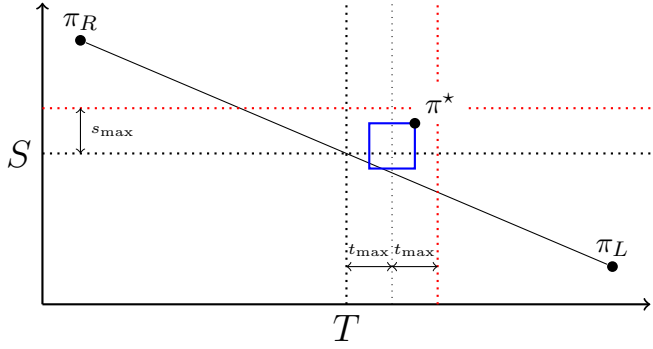
\begin{figure}
 \centering
	\tikzstyle{pre}=[<-,shorten <=1pt,>=stealth',semithick]
	\tikzstyle{post}=[->,shorten >=1pt,>=stealth',semithick]
	\tikzstyle{pline}=[very thin, color=gray]
	\tikzstyle{label}=[font=\Large]
	\tikzstyle{slabel}=[font=\small]
	\tikzstyle{bullet}=[fill, circle, inner sep=0.5mm]
	\tikzstyle{bline}=[dotted, thick, color=red]
	\tikzstyle{iline}=[dotted]
	\tikzstyle{arrow}=[<->]
	\tikzstyle{arrcap}=[auto,font=\small]
	\begin{tikzpicture}
		\draw[->,thick] (0,0) -- node[label, anchor=east] {$S$} (0,4) ;
		\draw[->,thick] (0,0) -- node[label, anchor=north] {$T$} (8,0);
		\node (piR) at (0.5, 3.5) [bullet] {};
		\node (piL) at (7.5, 0.5) [bullet] {}
			edge []	(piR);
		\node at (piR) [anchor=south] {$\pi_R$};
		\node at (piL) [anchor=south] {$\pi_L$};

		\draw[dotted, thick] (4,0) -- (4,4);
		
		\node (pt) at (intersection of 4,0--4,4 and 0.5,3.5--7.5,0.5) {};

		\draw[iline, thick] (pt -| 0,0) -- (pt -| 8,0);
		\draw[bline] ([yshift=+6mm]pt -| 0,0) -- ([yshift=+6mm]pt -| 8,0);
		\draw[iline] ([xshift=+6mm]4,0) -- ([xshift=+6mm]4,4);
		\draw[bline] ([xshift=+12mm]4,0) -- ([xshift=+12mm]4,4);

		\node (piS) at ([xshift=9mm,yshift=4mm]pt) {};
		\draw[color=blue,thick] (piS) rectangle ([xshift=-6mm, yshift=-6mm]piS);
		\node at (piS) [anchor=south west,fill=white] {$\pi^\star$};
		\node at (piS) [bullet] {};

		\draw[arrow] (4,0.5) -- node [arrcap] {\tiny{$\tmax{}$}} +(6mm,0);
		\draw[arrow] ([xshift=6mm]4,0.5) -- node [arrcap] {\tiny{$\tmax{}$}} +(6mm,0);
		\draw[arrow] (pt -| 0.5,0) -- node [arrcap, anchor=west] {\tiny{$\smax{}$}} +(0,6mm);
	\end{tikzpicture}
    \caption{Geometrical interpretation of Lemmas~\ref{lmm:paths-lagrange} and \ref{lmm:paths-step}.
    Paths are represented as points in the time-space coordinates.
    Path $\pi^\star$ is obtained by path-swapping paths $\pi_L$ and $\pi_R$. The blue rectangle is guaranteed by Lemma~\ref{lmm:paths-lagrange} to intersect with the segment from $\pi_L$ to $\pi_R$, while Lemma~\ref{lmm:paths-step} guarantees that there is at least one path-swapped solution having time coordinates between $t$ and $t + \tmax{}$ for any $t \in [t(\pi_R), t(\pi_L)]$, in this case $[T + \tmax{}, T + 2 \tmax{}]$.}
    \label{fig:phi-lambda2}
\end{figure}

Figure~\ref{fig:phi-lambda2} gives a geometrical interpretation of this lemmas and shows, in an intuitive way, that it is possible to path-swap the optimal basis $(\pi_L, \pi_R)$ computed by the cutting-plane algorithm (Lemma~\ref{lem:cp-complexity}) to get an additive $(\smax{}, 2\,\tmax{})$-approximation to the {\tt WCSPP} by carefully picking a swapping point $v$. This is a key result deployed to prove the following.

\deferred{path-swap-box}{Lemma}{lmm:path-swap-box}{
 Given an optimal basis $(\pi_L, \pi_R)$ with $t(\pi_L) > T$ and $t(\pi_R) < T$, there exists a swapping point $v^\star$ and a path-swapped path $\pi^\star = \paths{\pi_1}{\pi_2}{v^\star}$ such that $t(\pi^\star) \leq T + 2\,\tmax{}$ and $s(\pi^\star) \leq \varphi^\star + \smax{}$.
}
\begin{proof}
 Since $\paths{\pi_L}{\pi_R}{v_1} = \pi_R$ and $\paths{\pi_L}{\pi_R}{v_{n}} = \pi_L$, Lemma~\ref{lmm:paths-step} implies that there must exist some $v^\star$ such that the path $\pi^\star = \paths{\pi_L}{\pi_R}{v^\star}$ has time $t(\pi^\star) \in [ \, T + \tmax{} \,,\, T + 2\,\tmax{} \, ]$. Due to Lemma~\ref{lmm:paths-lagrange}, there are $s \geq s(\pi^\star) - \smax{}$ and $t \geq T$ (since $t + \tmax{} \geq t(\pi^\star) \geq T + \tmax{}$) such that $s + \lambda (t - T) = \varphi^\star$; hence $s \leq \varphi^\star$, which ultimately yields that $s(\pi^\star) \leq \varphi^\star + \smax{}$.
\end{proof}
\medskip The gap-closing procedure consists thus on choosing the best path-swap of the optimal basis $(\pi_L, \pi_R)$ with time-weight within $T + 2\,\tmax{}$.
The solution can be selected by scanning  left-to-right all the swapping points, and evaluating the time cost and space weight for each candidate.
This procedure can be implemented by keeping the time and space of the current prefix of $\pi_L$ and suffix of $\pi_R$, and by updating them every time a new swapping point is considered. Since each update can be performed in $O(1)$ time, we obtain the following lemma, which combined with Lemma~\ref{lem:cp-complexity}, proves our main Theorem~\ref{thm:main}.

\begin{Lemma}
 \label{lmm:ps-complexity}
 Given an optimal basis $(\pi_L, \pi_R)$ of problem $D'$, an additive $(\smax{}, 2\,\tmax{})$-ap\-pro\-xima\-tion to {\tt WCSPP} can be found in $O(n)$ time and $O(1)$ auxiliary space.
\end{Lemma}

\section{Experimental results}
\label{sec:experimental}

We describe here the preliminary results we obtained by executing \COMPNAME{}, an in-memory \texttt{C++} implementation of our \LZSE{}-based data-compression scheme introduced in this paper. These experiments aim not only at establishing the ultimate performance of our compressor, but also at investigating the following three issues:

\noindent {\bf 1) Trade-off range} In Section~\ref{sec:lzss-parse-modeling} we motivated the interest in the Time-Constrained Space-Optimal \LZSE{}-Parsing problem by showing a series of pathological texts for which the \LZSE{}-parsings exhibit wide space-time trade-offs. In this section, we provide experimental evidence that these pathological cases do occur in practice, so that the design of a flexible compressor, as the one we propose in this paper, is worth not only in theory!

\noindent {\bf 2) Estimating compression ratio} The number of phrases is a popular metric for estimating the compression ratio induced by a \LZSE{}-parsing. Ferragina et al.\@ showed \citep{FNV} that this is a simplistic metric, since there is a $\Omega\left(\frac{\log \log n}{\log n}\right)$ multiplicative gap in the compressed-space achieve by the bit-optimal parsing and the greedy one. In this section we deepen this argument by comparing experimentally the time-space trade-off when compressed space is either estimated exactly or approximated by the number of phrases in the parsing. The net result is to show that the number-of-phrases is a bad estimate for the space-occupancy of a \LZSE{}-based compressor, so the space-time trade-offs obtained by algorithms based only on this measure can be widely off-mark of the true ones.

\noindent {\bf 3) Comparing to the state-of-the-art} Our experiments are executed over datasets of several types and against many state-of-the-art compression libraries. 
We executed the experiments over datasets of several types of data:
\emph{\texttt{Wikipedia}}\footnote{Downloaded from \url{http://download.wikimedia.org/enwiki/latest/enwiki-latest-pages-articles.xml.bz2}} (natural language), 
\emph{\texttt{DBLP}}\footnote{Downloaded from \url{http://dblp.uni-trier.de/xml/}} (XML), 
\emph{\texttt{PFAM}} (biological data, \citep{PFAM}), and
\emph{\texttt{U.S.~Census}}\footnote{Downloaded from \url{http://www2.census.gov/census_2000/datasets/Summary_File_1/0Final_National/all_0Final_National.zip}} (database). 
Each dataset consists of a chunk of 1GB. We compared our compressor \COMPNAME{} against the most popular and top-performing compressors belonging to the two main families: \LZSE{}-based and BWT-based.
From the former family, we included:
\begin{inparaenum}[(i)]
\item \emph{\texttt{zlib}} which is the core of the well-known \texttt{gzip} compressor;
\item \emph{\texttt{LZMA2}} which is the core of \texttt{7zip} compressor and is appreciated for its high compression ratio and competitive decompression speed. 
\end{inparaenum}
From the latter family, we included:
\begin{inparaenum}[(i)]
\item \emph{\texttt{bzip2}} which is a general purpose compressor available on any Linux distributions; and
\item \emph{\texttt{BWT-Booster}} which is the state-of-the-art for BWT-based compressors \citep{FGMS05}.
\end{inparaenum}
Moreover, we included \emph{\texttt{Snappy}} and \LZF{} which are highly engineered \LZSE{}-compressors used in BigTable \citep{BigTable} and Hadoop, offering an amazingly fast decompression speed.

Each decompressor has been implemented in \verb=C++=, to work in-memory, and it has been compiled with \verb=g++= version \verb=4.6.3= with options \verb$-O3$ \verb$-fomit-frame-pointer$ \verb$-march=native$ and evaluated on a machine with the following characteristics:
\begin{inparaenum}[(i)]
\item processor: Intel Core 2 Duo P8600, with 64k of \verb=L1= cache and 3mb of \verb=L2= cache;
\item RAM: 4GB \verb=DDR3= \verb=PC3-8500=;
\item Operating system: Ubuntu 12.04.
\end{inparaenum}

\paragraph{Implementation details}
In implementing \COMPNAME{} we resorted to a simple byte-oriented encoder for the \LZSE{}-phrases which alleviates the detrimental effects of branching codes. Encoding a phrase requires at most $2$ bytes for the length and $4$ bytes for the distance (so $\smax = 48$ bits). Then we modeled the time-costs of the edges by three values which have been determined through many benchmarks (details in the journal paper), and we got $\tmax \approx 0.125 \mu s$.

In order to create the graph $\G{}$ we need to derive the space-costs and the time-weights of its edges. For the former, we have defined and implemented two different models which take into account either the overall compressed space in bits (the ``full'' model) or, the simplistic model which uses just the number of phrases constituting a parsing (the ``fixed'' model). After numerous experiments (details in the journal paper), the ``full'' model is constructed in such a way that the decoding cost of a phrase \LZP{d}{\ell} is \miss{1} if $d < 16000$, \miss{2} if $d \in [16000, 2300000]$, and \miss{3} otherwise, where the parameters $\miss{i}$ are derived by executing a proper benchmark over our machine (details in the journal paper).

At compression time, the user can specify a \emph{time bound} $T$ (in millisecs) or a \emph{compression level} $C = (T - T_t) / (T_s - T_t)$, where $T_t$ is the decompression time of the time-optimal parsing and $T_s$ is the decompression time of the most succinct space-optimal parsing.
We notice that compression level $C=0$ (resp. $C=1$) corresponds to the parsing with fastest decompression time (resp. smallest compressed space).

\begin{figure*}
\centering
\begin{minipage}{0.94\textwidth}
  \begin{minipage}{1\textwidth}
    \subfloat{ \includegraphics[width=0.50\textwidth]{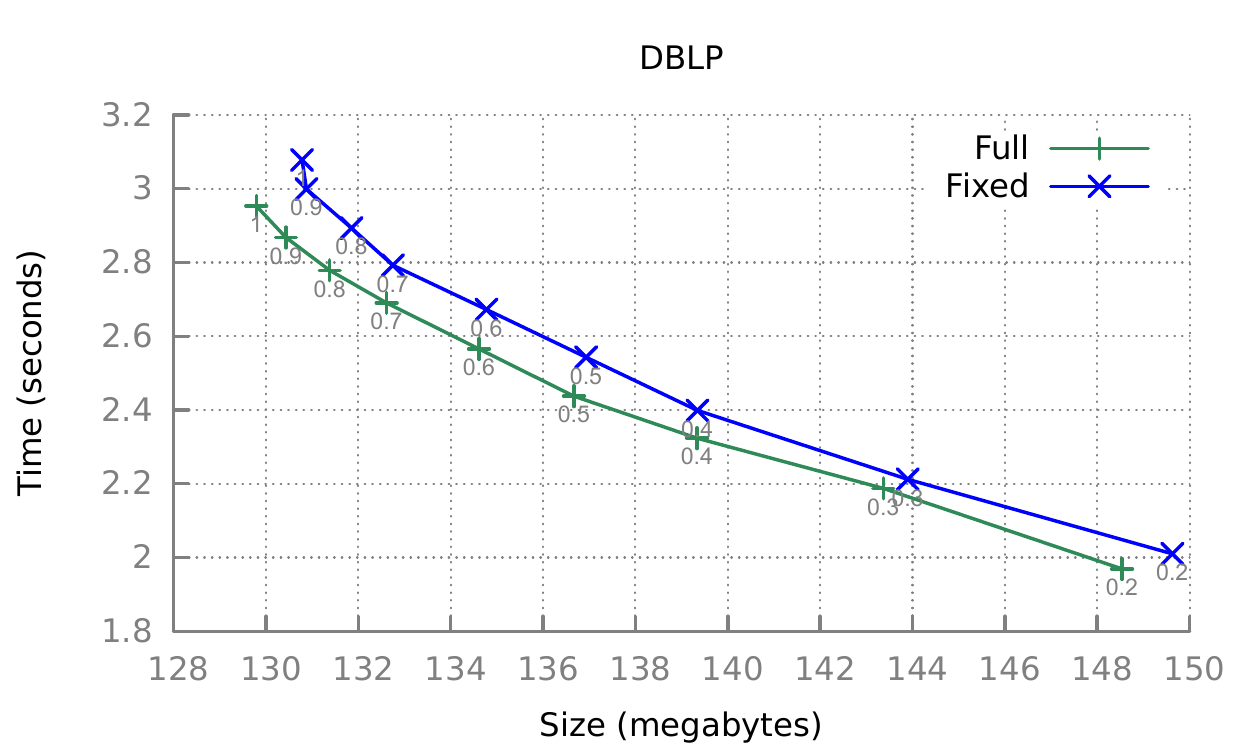} \includegraphics[width=0.50\textwidth]{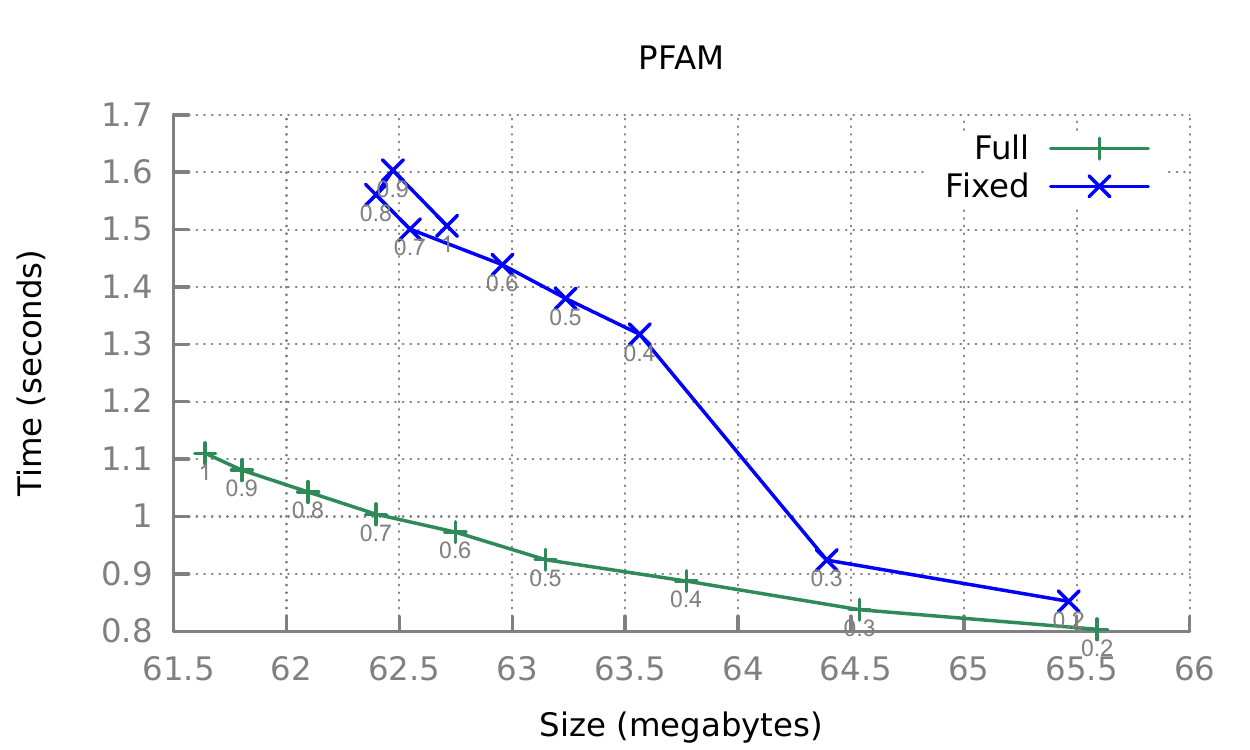}
    }\\
    \subfloat{
       \includegraphics[width=0.50\textwidth]{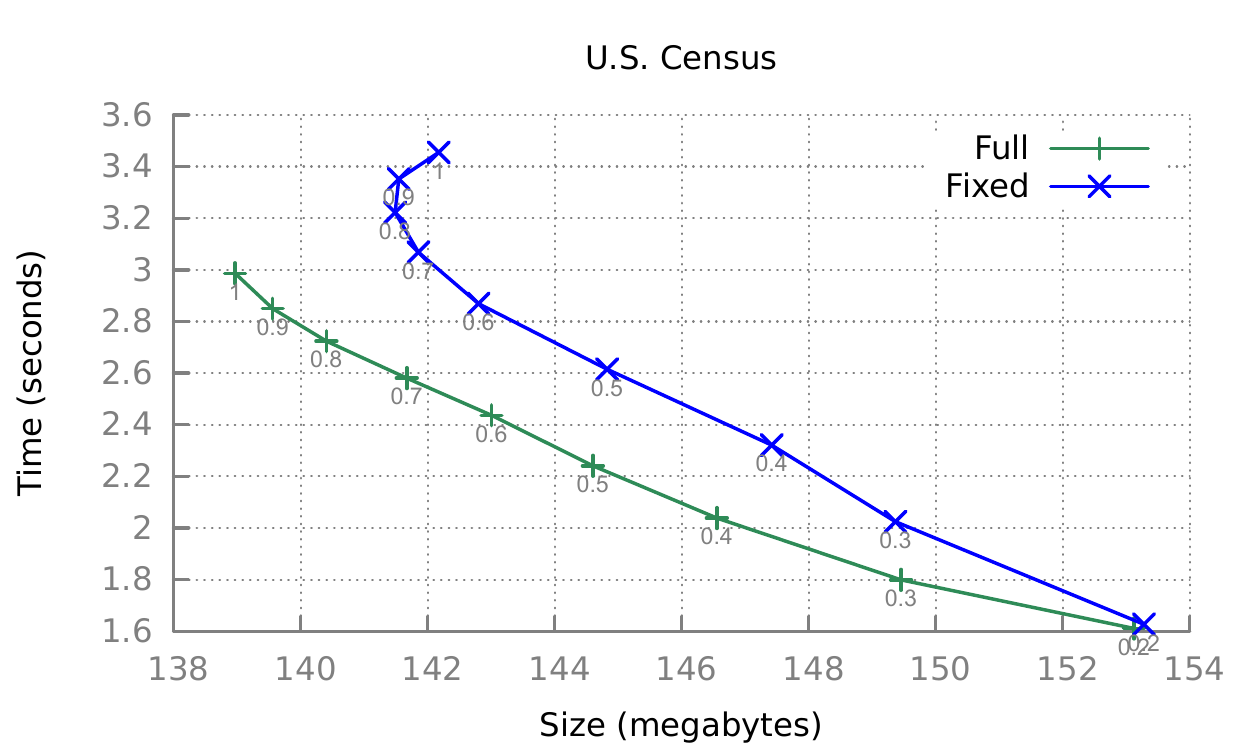}
       \includegraphics[width=0.50\textwidth]{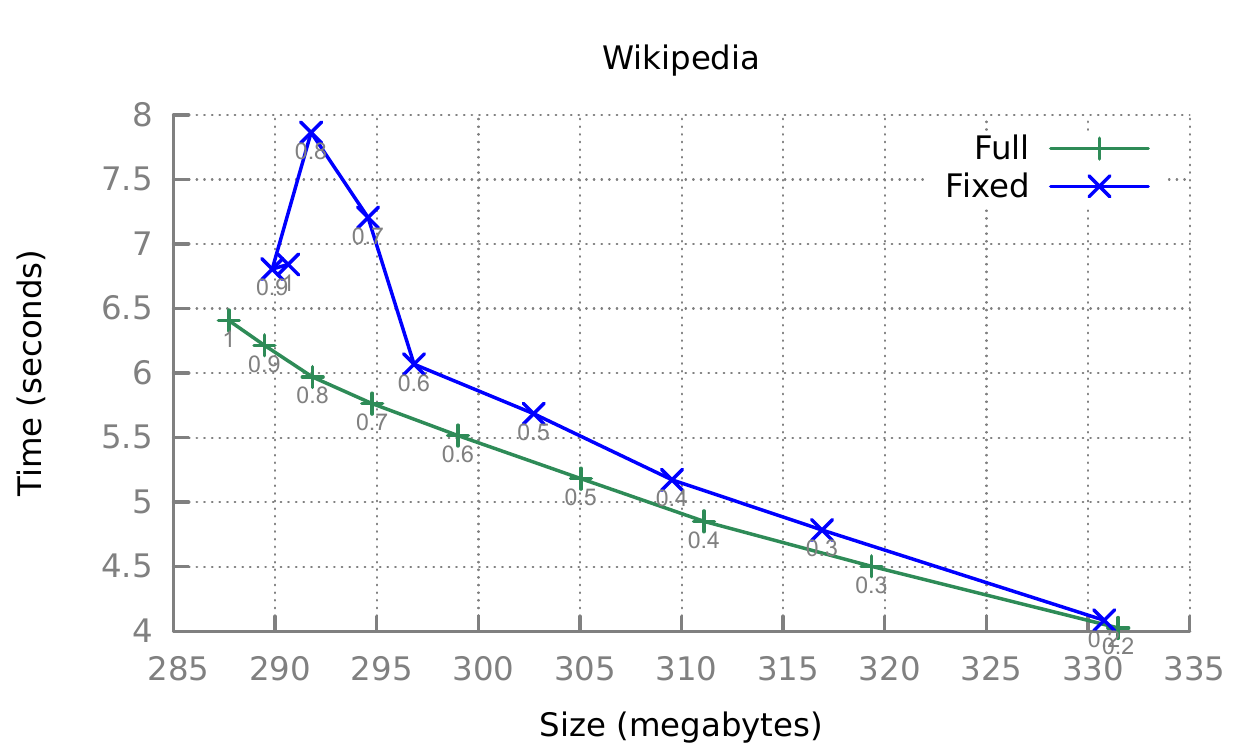}
    }\\
    \caption{\label{fig:test-fixfull}Decompression time and compressed space trade-offs obtained by changing the compression level from $0.2$ to $1$. Every point in each curve corresponds to a parsing whose compression level is indicated close to the point itself. In the ``full'' model (green line), the space cost of a codeword is given by its length, while in the ``fixed'' model the space cost is unitary.}
  \end{minipage}\\
  \begin{minipage}{1\textwidth}
    \renewcommand{\thempfootnote}{\fnsymbol{mpfootnote}}
    \resizebox{1\columnwidth}{!}{
\subfloat{ \begin{tabular}{c l r r} \toprule \multicolumn{1}{c}{Dataset} & \multicolumn{1}{c}{Parsing} & \multicolumn{1}{c}{Compressed size} & \multicolumn{1}{c}{Decompression time} \\ & & \multicolumn{1}{c}{(MB)} & \multicolumn{1}{c}{(seconds)}\\ \toprule
\multirow{15}{*}{\texttt{DBLP}}
& \COMPNAME{} - 1 & 129.8 & 2.95 \\
& \COMPNAME{} - 0.9 & 130.4 & 2.86 \\
& \COMPNAME{} - 0.8 & 131.4 & 2.77 \\
& \COMPNAME{} - 0.7 & 132.6 & 2.69 \\
& \COMPNAME{} - 0.6 & 134.6 & 2.56 \\
& \COMPNAME{} - 0.5 & 136.7 & 2.43 \\
& \COMPNAME{} - 0.4 & 139.3 & 2.32 \\
& \COMPNAME{} - 0.3 & 143.4 & 2.18 \\
& {\bf \COMPNAME{} - 0.2} & {\bf 148.5} & {\bf 1.96} \\
\cmidrule(r){2-4}
& \texttt{Snappy} & 323.4 & 2.13\\
& \texttt{LZ4} & 214.7 & 1.98\\
& \texttt{zlib} & 190.5 & 11.65\\
& \texttt{LZMA2} & 186.6 & 20.47\\
& \texttt{bzip2} & 121.4 & 48.98\\
& \texttt{BWT-Booster} & 98.2 & $> 100$\\
\midrule
\multirow{15}{*}{\texttt{PFAM}}
& {\bf \COMPNAME{} - 1} & {\bf 61.6} & {\bf 1.11} \\
& \COMPNAME{} - 0.9 & 61.8 & 1.08 \\
& \COMPNAME{} - 0.8 & 62.1 & 1.04 \\
& \COMPNAME{} - 0.7 & 62.4 & 1.00 \\
& \COMPNAME{} - 0.6 & 62.7 & 0.97 \\
& \COMPNAME{} - 0.5 & 63.1 & 0.92 \\
& \COMPNAME{} - 0.4 & 63.8 & 0.88 \\
& \COMPNAME{} - 0.3 & 64.5 & 0.83 \\
& \COMPNAME{} - 0.2 & 65.6 & 0.80 \\
\cmidrule(r){2-4}
& \texttt{Snappy} & 147.6 & 1.70\\
& \texttt{LZ4} & 74.4 & 1.41\\
& \texttt{zlib} & 62.3 & 7.63\\
& \texttt{LZMA2} & 49.5 & 7.16\\
& \texttt{bzip2} & 48.7 & 21.65\\
& \texttt{BWT-Booster} & 54.7 & $> 100$\\
\bottomrule \end{tabular} }
\subfloat{ \begin{tabular}{c l r r} \toprule \multicolumn{1}{c}{Dataset} & \multicolumn{1}{c}{Parsing} & \multicolumn{1}{c}{Compressed size} & \multicolumn{1}{c}{Decompression time} \\ & & \multicolumn{1}{c}{(MB)} & \multicolumn{1}{c}{(seconds)}\\ \toprule
\multirow{15}{*}{\texttt{U.S.\@ Census}}
& \COMPNAME{} - 1 & 139.0 & 2.98 \\
& \COMPNAME{} - 0.9 & 139.6 & 2.84 \\
& \COMPNAME{} - 0.8 & 140.4 & 2.72 \\
& \COMPNAME{} - 0.7 & 141.7 & 2.58 \\
& \COMPNAME{} - 0.6 & 143.0 & 2.43 \\
& \COMPNAME{} - 0.5 & 144.6 & 2.24 \\
& {\bf \COMPNAME{} - 0.4} & {\bf 146.6} & {\bf 2.03} \\
& \COMPNAME{} - 0.3 & 149.5 & 1.79 \\
& \COMPNAME{} - 0.2 & 153.1 & 1.61 \\
\cmidrule(r){2-4}
& \texttt{Snappy} & 324.1 & 2.28\\
& \texttt{LZ4} & 225.0 & 2.01\\
& \texttt{zlib} & 176.4 & 11.44\\
& \texttt{LZMA2} & 174.7 & 20.34\\
& \texttt{bzip2} & 180.7 & 50.40\\
& \texttt{BWT-Booster} & 141.9 & $> 100$\\
\midrule
\multirow{15}{*}{\texttt{Wikipedia}}
& \COMPNAME{} - 1 & 287.7 & 6.40 \\
& \COMPNAME{} - 0.9 & 289.5 & 6.21 \\
& \COMPNAME{} - 0.8 & 291.8 & 5.96 \\
& \COMPNAME{} - 0.7 & 294.8 & 5.76 \\
& \COMPNAME{} - 0.6 & 299.0 & 5.51 \\
& \COMPNAME{} - 0.5 & 305.0 & 5.18 \\
& \COMPNAME{} - 0.4 & 311.1 & 4.85 \\
& \COMPNAME{} - 0.3 & 319.3 & 4.50 \\
& {\bf \COMPNAME{} - 0.2} & {\bf 331.5} & {\bf 4.02} \\
\cmidrule(r){2-4}
& \texttt{Snappy} & 585.7 & 2.84\\
& \texttt{LZ4} & 435.1 & 2.63\\
& \texttt{zlib} & 380.5 & 17.63\\
& \texttt{LZMA2} & 363.5 & 39.09\\
& \texttt{bzip2} & 304.5 & 66.64\\
& \texttt{BWT-Booster} & 228.8 & $> 100$\\
\bottomrule \end{tabular} }

    }
\caption{Rows ``\COMPNAME{} - $c$'' stands for the performance of our implementation of the Time-Constrained Space-Optimal \LZSE{} parsing with compression level $c$.
}
    \label{tab:dectimes}
  \end{minipage}
\end{minipage}
\end{figure*}

\paragraph{Experimental trade-off spectrum}
We observed the experimental shape of the time-space trade-offs curve by compressing each dataset with linearly varying compression level from $0.2$ to $1$, considering both the fixed and the full model. Figure~\ref{fig:test-fixfull} shows that, for the full model, the space-time trade-off curve of the linearly changing compression level is actually linear too, which clearly shows that the trade-off can be effectively controlled in a principled way by our compressor.

Results are instead far less significant for the fixed model, in which space is estimated as the number of phrases in the parsing. Even if its curve is close to the one of the full model for \texttt{DBLP}, the curves are significantly different for the other three datasets. Moreover, the space-optimal parsing generated in the fixed-model with compression level 1 (which is equivalent to the greedy parsing) is dominated by the parsings generated with compression levels from $0.7$ to $0.9$ in \texttt{U.S.\@ Census}, while parsings with compression level $0.7$ and $0.8$ are dominated by parsings with compression level $0.9$ and $1$ in \texttt{Wikipedia}. This clearly shows that the number of phrases is a poor metric for estimating the compression ratio of a parsing, and it offers a very simplistic estimate of the decompression time.

\paragraph{Comparison with state-of-the-art}
Table~\ref{tab:dectimes} reports the performance of the various compression algorithms on the datasets. Results show that the performance of our \COMPNAME{} are extremely good. On the one hand, it generates parsings with decompression time better than those of \LZF{} in three out of four datasets (\texttt{DBLP}, \texttt{PFAM}, \texttt{U.S.\@ Census}), whereas for the fourth dataset (\texttt{Wikipedia}) \COMPNAME{} achieves a decompression time which is a little bit worse than \LZF{} but with a significantly improved compression ratio. On the other hand, its compression ratio at higher compression levels is close to the best one, namely that of \texttt{LZMA2} (excluding \texttt{BWT-Booster}, which exhibit an exceedingly slow decompression time), but with an order of magnitude faster decompression speed. Compression ratios of \COMPNAME{} are indeed very remarkable, because it uses a very simple byte-oriented encoder opposed to the statistical Markov-Chain encoder used in \texttt{LZMA2}.

Overall, these results show that not only our approach allows to effectively control the time-space trade-off in a practical yet principled manner; by explicitly taking into account both decompression-time and compressed-space, \COMPNAME{} leads to parsings which are \emph{faster to decode and more space-succinct} than those generated by highly tuned and engineered parsing \emph{heuristics}, like those of \texttt{Snappy} and \LZF{}.
\section{Conclusions}
\label{sec:conclusions}

We conclude this paper by mentioning two interesting future directions where the novel optimization-based approach proposed in this paper could be tested. One, whose practical impact that could hardly be overestimated, concerns the practical impact of these techniques on real big-data applications and their storage systems, like Hadoop. The second question, more of a theoretical vein, is whether it is possible to extend this novel bicriteria optimization approach to other interesting compressor families such as {\tt PPM} and {\tt BWT}.

\newpage
\bibliographystyle{plain}
\bibliography{pruning}

\newpage

\appendix

\section{Proof of Theorem~\ref{thm:multiplicative-apx}}
\label{sec:multiplicative-apx}

\recall{multiplicative-apx}
\begin{proof}
The main idea behind this theorem is that of solving {\tt WCSPP} with the additive-approxi\-ma\-tion algorithm of Theorem~\ref{thm:main}, thus obtaining a path $\pi^\star$. Then, we check whether $\frac{1}{\epsilon}\, \smax{} \leq \varphi^\star$ and $\frac{1}{\epsilon}\,\tmax{} \leq T$ and, in this case, we return $\pi^\star$ which is guaranteed to satisfy $s(\pi^\star) \leq \varphi^\star + \smax{} \leq (1+\epsilon)\varphi^\star$ and $t(\pi^\star) \leq T + 2\tmax{} \leq (1+2\epsilon)T$, thus being a $\left(\epsilon,2\,\epsilon\right)$-multiplicative approximated solution for {\tt WCSPP}. Recall that $\varphi^\star$ is known from the solution of the Dual Lagrangian.

Otherwise, we execute an exhaustive-search algorithm which takes sub-quadratic time because it explores a very small space of candidate solutions because either $\frac{1}{\epsilon}\, \smax{} > \varphi^\star$ or $\frac{1}{\epsilon}\,\tmax{} > T$. Consider for instance the case $\frac{1}{\epsilon} \; \smax{} > \varphi^\star$ (the other case is symmetric, and leads to the same conclusion): we can prove that we can find the optimal path by enumerating a small set of paths in \G{} via a breadth-first visit delimited by a pruning condition.

The key idea (details in the full paper) is to prune a sub-path $\pi'$, going from node $1$ to some node $v'$, if $s(\pi') > \frac{1}{\epsilon}\, \smax{}$; this should be a relatively easy condition to satisfy, since $\smax{}$ is the maximum space-cost of one single edge (hence of an \LZSE{}-phrase). If this condition holds, then $s(\pi') > \frac{1}{\epsilon}\, \smax{} > \varphi^\star$ (see before) and so $\pi'$ cannot be optimal and thus must be discarded. We can also prune $\pi'$ upon finding a path $\pi''$ which arrives to a farther node $v'' > v'$ while requiring the same compressed-space and decompression-time of $\pi'$ (Lemma~\ref{lmm:path-suffix}).

So all paths $\pi$ which are not pruned guarantee that $s(\pi) \leq \frac{1}{\epsilon}\, \smax{}$ and thus $t(\pi) \leq s(\pi) \tmax{} \leq \frac{1}{\epsilon}\, \smax{}\tmax{}$ (just observe that every edge has integral time-weight in the range $[1, \tmax{}]$). Therefore we can adopt a sort of dynamic-programming approach, \'a la knapsack, which computes the exact optimal solution (not just an approximation) by filling a bi-dimensional matrix $m$ of size $S \times U$, where $S = \frac{1}{\epsilon}\, \smax{}$ is the maximum space-cost admitted and $U = S\,\tmax{}$ is the maximum time-cost admitted for a candidate solution/path. Entry $m[s,t]$ stores the farthest node in \G{} reachable from $1$ by a path $\pi$ with $s=s(\pi)$ and $t=t(\pi)$. These entries are filled in $L$ rounds, where $L\leq \frac{1}{\epsilon}\, \smax{}$ is the maximum length of the optimal path (just observe that every edge has integral space-weight in the range $[1, \smax{}]$). Each round $\ell$ constructs the set $X_\ell$ of paths having length $\ell$ and starting from node $1$ which are candidate to be the optimal path. These paths are generated by extending the paths in $X_{\ell - 1}$, via the visit of the forward star of their last nodes, in $O(n \log n)$ time and $O(n)$ space according to the {\tt FSG}-algorithm \cite{FNV}. Each generated path $\pi'$ is checked for usefulness: if $s(\pi') > \frac{1}{\epsilon}\, \smax{}$, then $\pi'$ is pruned; it is also pruned if its last node is to the left of vertex $m[s(\pi'), t(\pi')]$. Otherwise, we set that node into that entry (Lemma~\ref{lmm:path-suffix}). The algorithm goes to the next round by setting $X_\ell$ as the paths remaining after the pruning.

As far as the time complexity of this process is concerned, we note that $|X_\ell| \leq S\, U = O\left(\frac{1}{\epsilon^2} \log^3 n\right)$. The forward-star of each vertex needed for $X_\ell$ can be generated by creating the pruned $\widetilde{\G{}}$ in $O(n \log n)$ time and $O(n)$ space. Since we have a total of $L\leq \frac{1}{\epsilon}\, \smax{} = O(\frac{1}{\epsilon}\,\log n)$ rounds, the total time is $O\left(\frac{1}{\epsilon^3} \log^4 n + \frac{1}{\epsilon}\, n \log^2 n\right)$.

As a final remark, note that one could instead generate the forward star of all nodes in $\widetilde{\G{}}$, which requires $O(n \log n)$ time \emph{and} space, and then use them as needed. This would simplify and speed-up the algorithm achieving $O\left(\frac{1}{\epsilon} n \log n\right)$ total time; however, this would increase the working space to the super-linear $O(n \log n)$ in the size $n$ of the input string $\Txt$ to be compressed, which is best avoided in the context these algorithms are more likely to be used in.
\end{proof}

\end{document}